\documentclass[preprint2,10pt]{aastex}

\usepackage{amsmath}
\usepackage{natbib}
\usepackage{here}
\newcommand{\prep}[1]{#1}
\newcommand{\manu}[1]{}

\prep{\setlength\textwidth{19cm}}
\prep{\setlength\textheight{24.5cm}}
\prep{\setlength\columnsep{0.5cm}}
\prep{\setlength\topmargin{-2.5cm}}
\prep{\hoffset =-1.5cm }
\prep{\renewcommand{\baselinestretch}{0.92}} 

\newcommand{\mywidth}{8.3 cm}

\def\dop#1#2{{#1}\cdot{#2}}

\def\deg{^\circ}
\newcommand{\te}[1]{\times10^{#1}}

\newcommand{\bn}{\boldsymbol{n}}
\newcommand{\bw}{\boldsymbol{w}}
\newcommand{\bi}{\boldsymbol{i}}
\newcommand{\bj}{\boldsymbol{j}}
\newcommand{\bk}{\boldsymbol{k}}

\newcommand{\one}{$I$}
\newcommand{\two}{$I\!I$}
\newcommand{\three}{$I\!I\!I$}
\newcommand{\four}{$IV$}
\newcommand{\five}{$V$}
\newcommand{\six}{$V\!I$}

\def\m@th{\mathsurround=0pt}
\def\EQM#1{\vcenter{\normalbaselines\m@th
    \ialign{${\displaystyle ##}$\hfil&&\ ${\displaystyle ##}$\hfil\crcr
    \mathstrut\crcr\noalign{\kern-\baselineskip}
    \noalign{\smallskip}
    #1\crcr\mathstrut\crcr\noalign{\kern-\baselineskip}}}}

\newcommand{\be}{\begin{equation}}
\newcommand{\ee}{\end{equation}}

\newcommand{\bpm}{\begin{pmatrix}}
\newcommand{\epm}{\end{pmatrix}}

\newcommand\captiona{
 {\bf Uranus precession frequency in presence of a heavy
satellite.}
Uranus effective precession constant as a function of the distance of an
additional satellite of mass $m=0.01\ M_U$ ({\bf a}), $m=0.005\ M_U$
({\bf b}), and $m=0.001\ M_U$ ({\bf c}), where $M_U$ is the mass of
Uranus \citep{Boue_Laskar_Icarus_2006}.  For this calculation, the
semi-major axis and the eccentricity of Uranus are set to the current
values, and the satellite is assumed to have a circular orbit. All
inclinations as well as the obliquity are set to 0.  }

\newcommand\figa{
\begin{figure}[h!]
\begin{center}
\includegraphics[width=\mywidth]{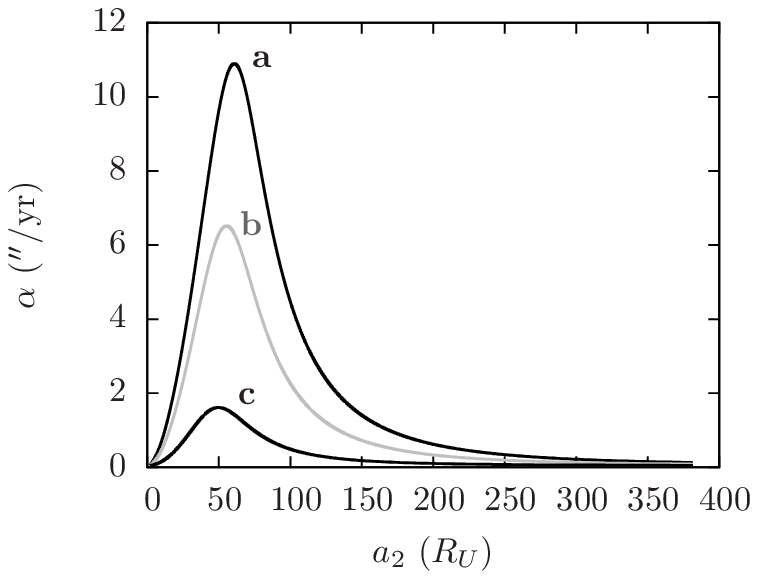}
\caption{\captiona}
\label{Figa}
\end{center}
\end{figure}
}

\newcommand\captionb{{\bf Comparison between orbital evolution and
obliquity increase.} {\bf a,b,} Example of orbital evolution of the
giant planets during the planetary migration over 2 Myr. ({\bf a})
semi-major axis, minimum, and maximum heliocentric distances. ({\bf b})
Uranus inclination. {\bf c,} Maximal tilt starting from zero obliquity
for any orbital evolution with the same semi-major axis, eccentricity
and inclination as Uranus in subfigures ({\bf a,b}). 
In this calculation (see Fig.~\ref{Figd}), the effect of an additional
satellite at 50 Uranian radii is implicitely taken into account in the
precession constant. We considered three different masses for the
satellite : $10^{-4}\ M_U$, $10^{-3}\ M_U$, and $0.01\ M_U$, where $M_U$
is the mass of Uranus. For each mass, the satellite eccentricity is set
to 0 (lower boundary) and 0.5 (upper boundary). {\bf d}, Results of a
numerical integration with a satellite of mass $m=0.01\ M_U$ (black
curve).  The satellite is ejected by a close encounter with Saturn at
$t=0.38$ Myr. Once the satellite is ejected, the obliquity remains
constant (gray curve). In these plots, the obliquity is measured
relative to the fixed plane orthogonal to the final total orbital
angular momentum.
}

\newcommand\figb{
\begin{figure}
\begin{center}
\includegraphics[width=\mywidth]{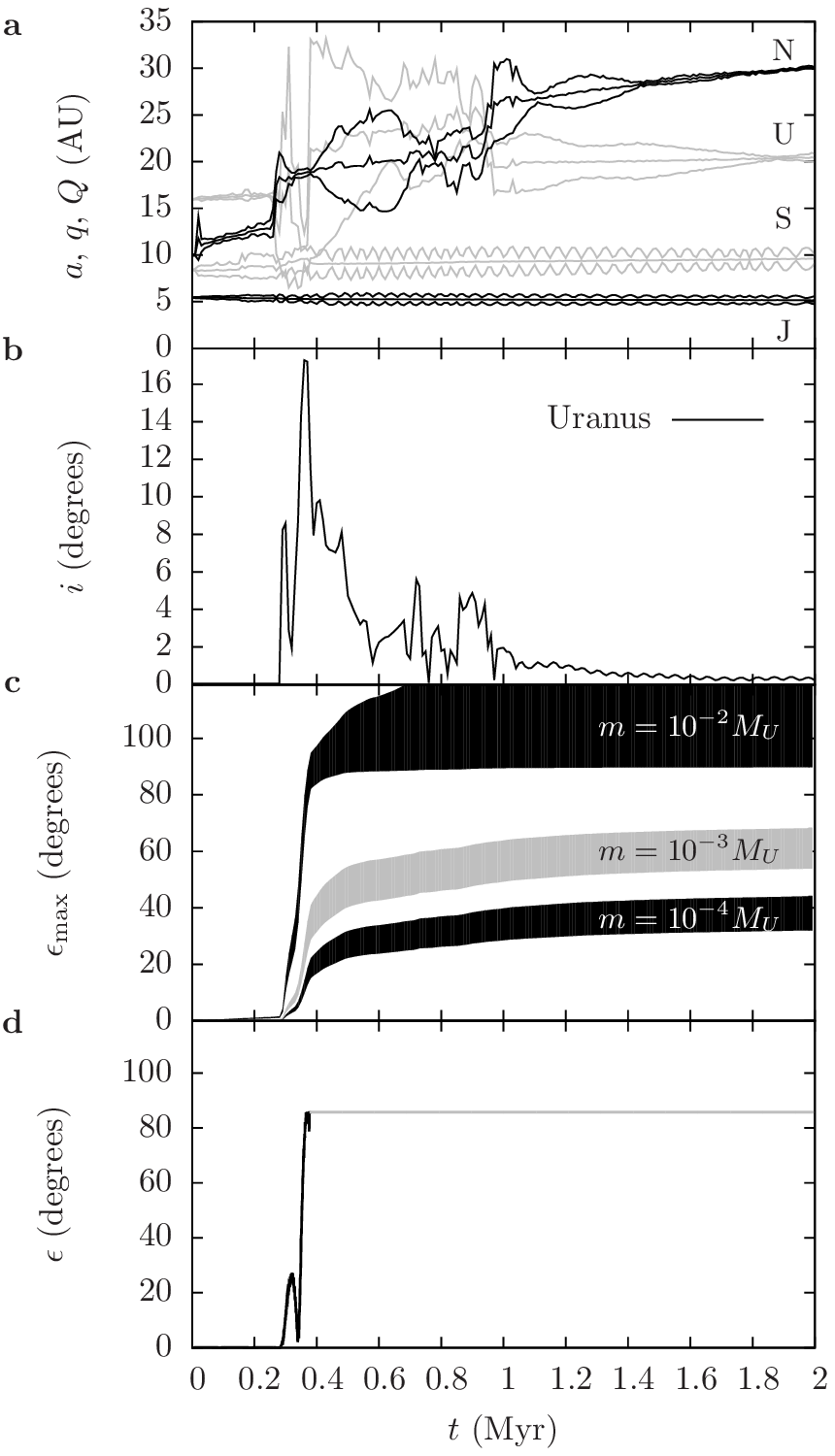}
\caption{\captionb}
\label{Figb}
\end{center}
\end{figure}
}

\newcommand\captionc{{\bf Distribution of Uranus final obliquity.} This
is the
result of $1\,700$ integrations of Uranus spin-axis with an additional
satellite : 100 per each of the 17 selected migrations. In black, the
cases where the satellite is ejected; in gray, the cases where the
satellite still orbits Uranus after 2 Myr (the end of the integrations
with a satellite). The first bin has been troncated for a better
visualization, its value is 644 (black) + 8 (gray). Among the
simulations with ejection of the satellite, there is a final obliquity
larger than $60\deg$ (resp. $90\deg$) in 220 cases (resp. 37 cases).
}

\newcommand\figc{
\begin{figure}[t]
\begin{center}
\includegraphics[width=\mywidth]{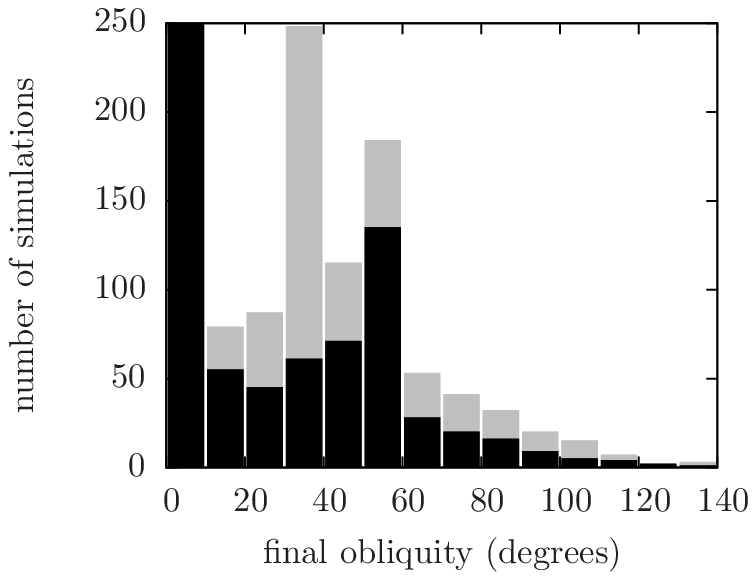}
\caption{\captionc}
\label{Figc}
\end{center}
\end{figure}
}

\newcommand\captiond{{\bf Constraints on the migration and on the mass
of the satellite.} {\bf a}, Lower boundary on the time required to tilt
Uranus by $97\deg$ as a function of the inclination of its orbit for
different precession constants \citep{Boue_etal_ApJL_2009}.  {\bf b},
Uranus' precession constant as a function of its semi-major axis for
different masses of the satellite in circular orbit
\citep{Boue_Laskar_Icarus_2006}.  In case of eccentric orbit of the
planet, the semi-major axis $a_U$ should be replaced by
$a_U\sqrt{1-e_U^2}$. For example, if during the planetary migration,
Uranus' inclination remains above $10\deg$ at least 1 Myr, then the
precession constant should be larger or equal to
$5^{\prime\prime}\!\!\cdot{\rm yr}^{-1}$ to tilt Uranus ({\bf a}).  Such
a precession constant can be reached with a satellite of mass
$m=0.001M_U$ if $a_U\sqrt{1-e_U^2}$ is less than $10$ AU during the
period of large inclination ({\bf b}).
}

\newcommand\figd{
\begin{figure}
\begin{center}
\includegraphics[width=\mywidth]{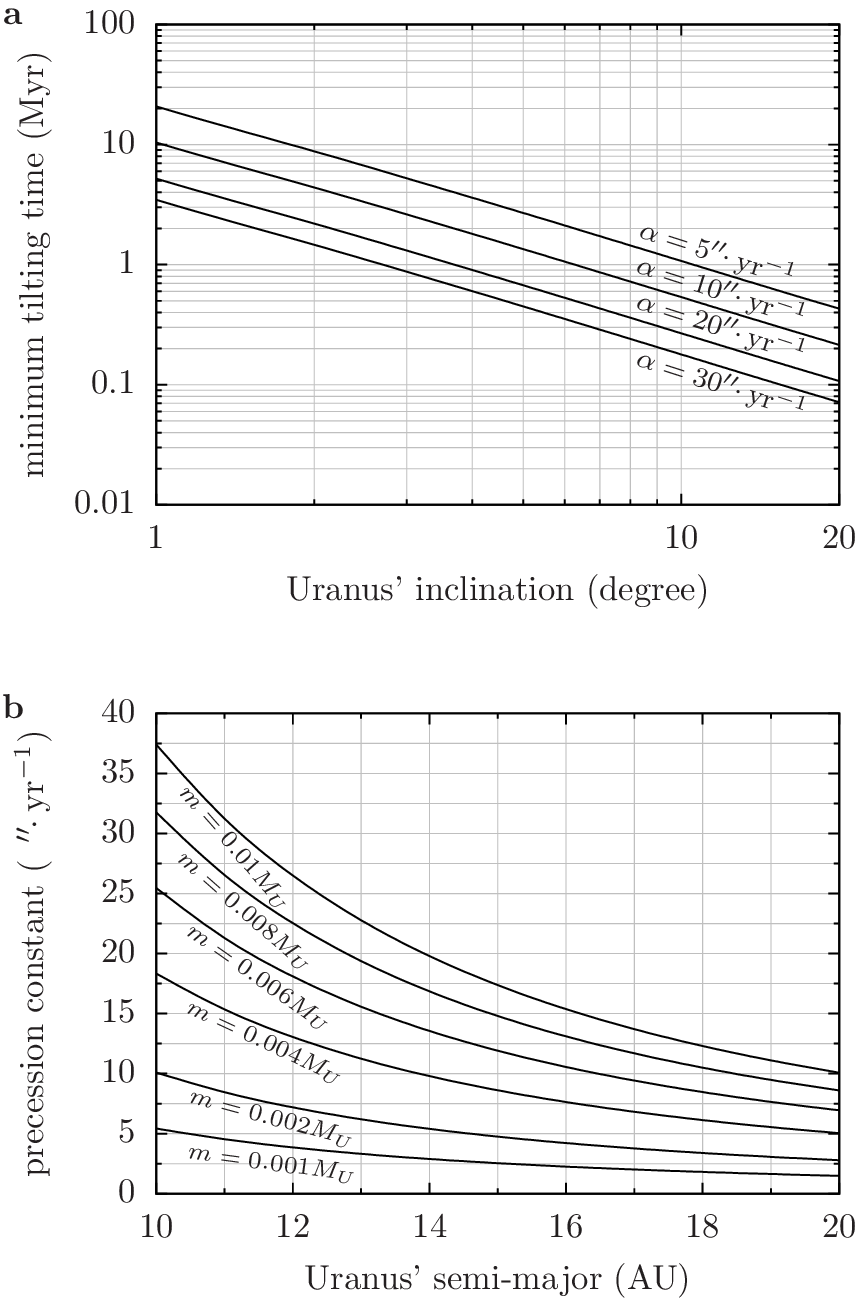}
\caption{\captiond}
\label{Figd}
\end{center}
\end{figure}
}

\newcommand\captione{
{\bf Details on Uranus tilting.} {\bf a}, Evolution of the resonant
angle $\psi = \phi_{\alpha}-\phi_{\nu}-\pi$ where $\phi_{\alpha}$ and
$\phi_{\nu}$ are angles measured positively from a reference direction
to the projection of Uranus' spin-axis $\bw$ and Uranus' orbit pole
$\bn$ into the $x$-$y$ plane, respectively. {\bf b}, Evolution of
Uranus' obliquity relative to the invariant plane. {\bf c}, Evolution of Uranus' orbital
inclination. {\bf d}, Evolution of Uranus' spin-axis.
The coordinates are $x=\sin\epsilon\cos\psi$, $y=\sin\epsilon\sin\psi$,
and $z=\cos\epsilon$. The thin black circles in the $x$-$y$ plane, that
correspond to the thin black lines in the $x$-$z$ plane and in the
$y$-$z$ plane, represent the locations where the obliquity is $45\deg$
(inner circle), and $90\deg$ (outer circle).
Uranus tilting is characterized by three resonant phases
labelled \two, and \four+\five{} separated by non-resonant evolutions
labelled \one, \three, and \six.
}

\newcommand\fige{
\begin{figure}[H]
\begin{center}
\includegraphics[clip=true,width=\mywidth]{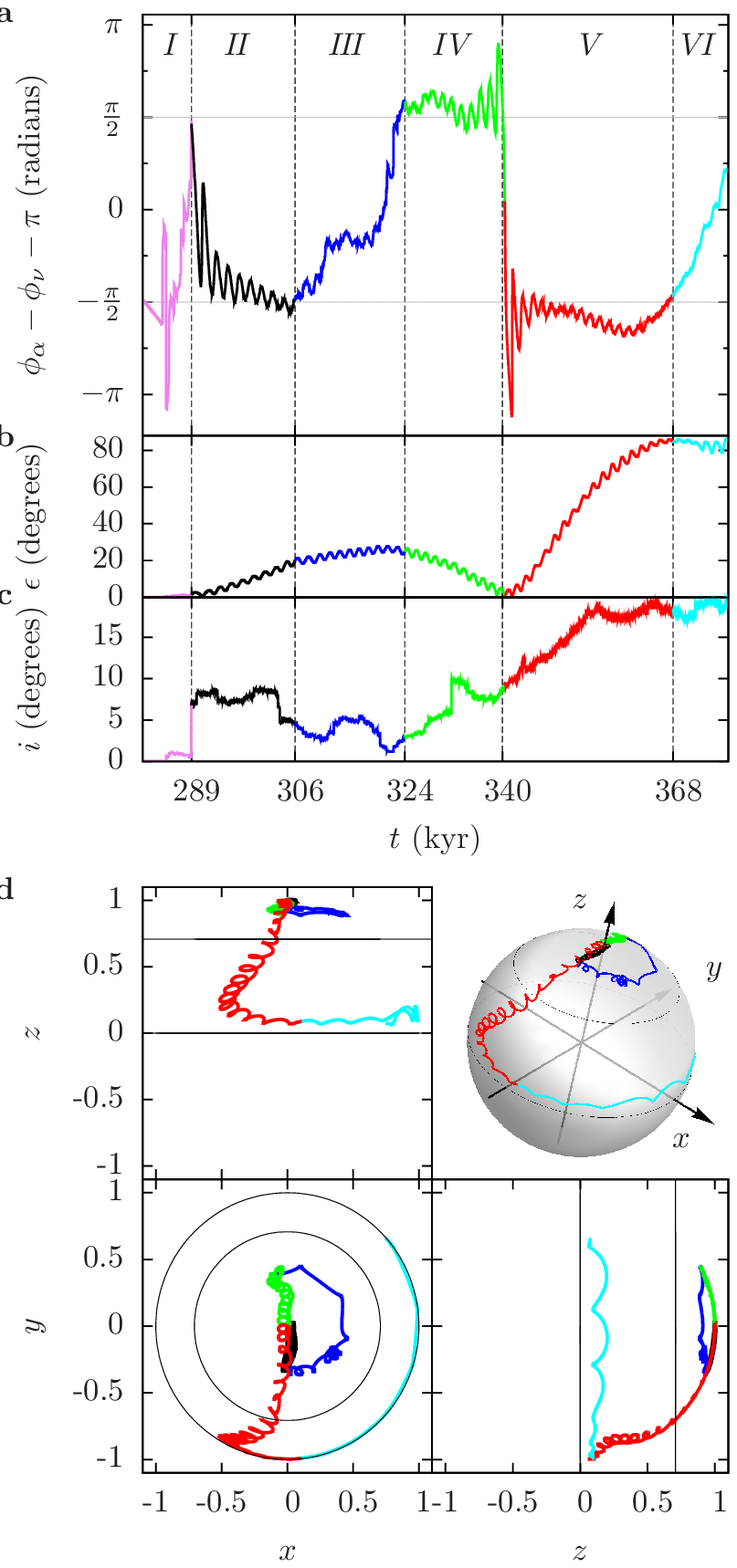}
\caption{\captione}
\label{Fige}
\end{center}
\end{figure}
}

\shorttitle{Uranus tilting}
\shortauthors{Bou\'e, \& Laskar}

\begin{document}
\slugcomment{{\sc \small The Astronomical Journal}}

\title{\prep{\uppercase}{A collisionless scenario for Uranus tilting}}

\author{\prep{\sc} Gwena\"el Bou\'e, and Jacques Laskar}
\affil{Astronomie et Syst\`emes Dynamiques, IMCCE-CNRS UMR8028,
Observatoire de Paris, UPMC, 77 Av. Denfert-Rochereau, 75014 Paris, France}
\email{boue@imcce.fr}

\date{\today}

\begin{abstract}
The origin of the high inclination of Uranus' spin-axis (Uranus'
obliquity) is one of the great unanswered questions about the Solar
system. Giant planets are believed to form with nearly zero obliquity,
and it has been shown that the present behaviour of Uranus' spin is
essentially stable. Several attempts were made in order to solve this
problem. Here we report numerical simulations showing that Uranus' axis
can be tilted during the planetary migration, without the need of a
giant impact, provided that the planet had an additional satellite and a
temporary large inclination. This might have happened during the giant
planet instability phase described in the Nice model. In our scenario,
the satellite is ejected after the tilt by a close encounter at the end
of the migration. This model can both explain Uranus' large obliquity
and bring new constraints on the planet orbital evolution.
\end{abstract}


\section{Introduction}
Today, Uranus' obliquity ($97\deg$) is essentially stable. This is due
to the regularity of its orbital motion and to the slow precession
motion of Uranus' axis compared to the secular frequencies of the Solar
system \citep{Laskar_Robutel_Nature_1993}.  It is sometimes believed
that a great collision with an Earth-sized protoplanet could be the
reason of Uranus large obliquity. But this straightforward scenario
hardly explains the presence of regular satellites orbiting Uranus in
its equatorial plane \citep{Korycansky_etal_Icarus_1990}.
However, the presence of satellites around a planet can increase its
precession rate of by a large amount depending on their mass and their
orbital parameters \citep{Tremaine_Icarus_1991, Goldreich_RvGeo_1966,
Ward_AJ_1975, Boue_Laskar_Icarus_2006}. For example, with a satellite of
mass $m=0.01M_U$, where $M_U$ is the mass of Uranus, the increase can
reach a factor 1000 (Fig.~\ref{Figa}). The maximal effect is obtained
for a satellite located at about 50 Uranian radii,
which is actually in the region where a satellite has been predicted by
some formation models \citep{Mosqueira_Estrada_Icarus_2003,
Mosqueira_Estrada_Icarus_2003b,
Estrada_Mosqueira_Icarus_2006}.  For
comparison, the most distant regular satellite of Uranus is Oberon, its
mass is $3.45\te{-5}M_U$, and its distance from Uranus' barycenter is 23
Uranian radii \citep{Laskar_Jacobson_AA_1987}.  The interactions between
spin-axes and secular motions of the planets are also strenghtened when
orbital inclinations are high.  Such conditions could be met during the
planetary migration.  Indeed, in the Nice scenario
\citep{Tsiganis_etal_Nature_2005}, Jupiter and/or Saturn should have
undergone close encounters with the ice giants to reach their present
eccentricities \citep{Morbidelli_etal_AA_2009}. These close encounters
can raise the inclinations.  Moreover, the additional satellite can be
ejected during one of these encounters. We therefore propose that Uranus
had an additional satellite and that its spin-axis was tilted during the
planetary migration.

\section{Numerical experiment}
The construction of such a scenario for Uranus tilting can be described
 in four steps.
\prep{\figa}
\prep{\figb}

First, we simulated the Nice model \citep{Tsiganis_etal_Nature_2005}. We
integrated $10\,000$ migrations of the giant planets over 10 Myr. For
these simulations,
\begin{eqnarray*}
\dot{a}&=&\frac{\delta a}{\tau}\exp{(-t/\tau)} \\
d{e}/dt&=&-e/(2\tau_e) \\
d{i}/dt&=&-i/(2\tau_i)
\end{eqnarray*}
with $\tau=2$ Myr, and $\tau_e=\tau_i=\tau/10$ as in
\citep{Lee_etal_Icarus_2007}. $a$, $e$ and $i$ are
respectively the semi-major axis, the eccentricity and the inclination
of the giant planets; $t$ is the time and $\delta a$ is the difference
between the initial semi-major axis of a planet and its current one.
The initial semi-major axis of Jupiter was set to 5.45 AU. The initial
semi-major axis of the other planets were obtained randomly with a
uniform distribution.  The initial semi-major axis of Saturn was varied
in the range 8.38-8.48 AU. The initial order of the ice giants was
modified compare to the current Solar system : the initial semi-major
axis of Neptune was varied in the range 9.9-12 AU and the initial
semi-major axis of Uranus was varied in the range 13.4-17.1 AU.  The
terrestrial planets are not taken into account in this study.

Then, out of the $5\,142$ simulations that survived without ejection or
planet collision, we selected those where the planet final order is the
same as in the Solar system. We obtained $1\,995$ different
integrations.  As the tilt requires high inclination, we kept only the
simulations where Uranus' inclination increases beyond a given
threshold. We set this threshold to $17\deg$ which limits the number of
simulations to 31.  Among these simulations, we rejected those where the
closest encounter between Uranus and any other planet is closer than 50
Uranian radii. With our criterion, we finally selected 17
simulations. One of them is displayed in
figure~\ref{Figb}a,~\ref{Figb}b. We call it the reference simulation.

In a third step, we computed the maximal effect of an additional
satellite on Uranus' obliquity in any orbital evolution with the same
semi-major axis, eccentricity and inclination as in the reference
simulation, regardless the conjugated angles. For that, we used the
expression of the effective precess\-ion constant as a function of the
satellite orbital parameters \citep{Boue_Laskar_Icarus_2006}.  Then, we
computed the maximal tilt given Uranus orbital evolution
\citep{Boue_etal_ApJL_2009}. In the calculations, the satellite is
at 50 Uranian radii in both circular and eccentric orbits.
Figure~\ref{Figb}c shows the maximal obliquity that has been reached in
these simulations. The evolution shows clearly that the tilt can only
occur when the inclination is high. In the present case, a satellite
with $m=0.01M_U$ is still necessary for the obliquity to reach $97\deg$.

Finally, we integrated the evolution of Uranus' spin-axis and the
additional satellite in the 17 selected simulations. Calculations of the
evolution of Uranus' spin-axis take into account the gravitational
torques exerted by the Sun, by the additional satellite and by all the
other giant planets. For each of the 17 planet migrations, we performed
100 integrations varying the initial semi-major axis of the satellite by
a small amount (15 meters). The final obliquity distribution is given in
figure~\ref{Figc}. In 644 cases, the obliquity does not exceed $10\deg$
because the satellite is ejected at the first encounter before the
increase of the inclinations. But, if the satellite survive the first
encounter, as in 62\% of the cases, then the obliquity can reach large
values. Among the integrations in which the satellite is ejected before
the end of the migration, there is a final obliquity larger than
$60\deg$ in 220 cases and an obliquity larger than $90\deg$ in 37 cases.

\prep{\figc}

\section{Dynamics of the tilt}
Here we explain the evolution of Uranus' spin-axis during the tilt
presented in the figure \ref{Figb}. The smooth evolution of the
obliquity
during the tilt (Fig.~\ref{Fige}{\bf b}) suggests that it is due
to a resonance. In the following, we show that the tilt actually occurs
during a $1\!\!:\!\!1$ spin-orbit resonance between the precession of
Uranus' axis and the regression of the node of its orbit.
The obliquity $\epsilon$ is measured relative to the
invariant plane orthogonal to the total orbital angular momentum at
the end of the simulation (10 Myr). Traditionally, it is defined
relative
to the orbital plane, but as the inclination rises to high values, it is
preferable to use the invariant plane in order to avoid artificial
evolution of the spin-axis. At the end of the reference simulation,
Uranus'
orbital inclination is very small ($0.0024\deg$) and the difference
between the two definitions is sufficiently small to be neglected.
Let $(\bi,\bj,\bk)$ be a base frame such that the $x$-$y$ plane
coincides with this invariant plane. We note $\bw$, Uranus' spin-axis,
and $\bn$, the normal of its orbit. The obliquity $\epsilon$ is thus
defined by $\cos\epsilon=\dop{\bk}{\bw}$. Let $\phi_\alpha$ and
$\phi_\nu$ be the angles measured positively from the reference direction
$\bi$ to the projections of $\bw$ and $\bn$ into the $x$-$y$ plane,
respectively. The evolution of $\psi=\phi_\alpha-\phi_\nu-\pi$ is not
steady but describes plateaus during the tilt (Fig.~\ref{Fige}{\bf a},
phases \two, and \four+\five). This confirms the $1\!\!:\!\!1$ resonance
between
the precession of Uranus' axis and the regression of the node of its
orbit.

In order to have a full understanding of the tilt, we now give the
equations of motion that will allow us to describe the spatial evolution
of the spin-axis displayed in the figure~\ref{Fige}{\bf d}.
\prep{\fige}
Let $\alpha$ be
Uranus' precession constant including the effect of the satellites,
and $\nu$ the regression frequency of Uranus' orbital node. In the
frame rotating around $\bk$ with the regression frequency $\nu$, the
Hamiltonian describing the spin-axis evolution reads
\be
H = -\frac{\alpha}{2}(\dop{\bn}{\bw})^2-\nu(\dop{\bk}{\bw}).
\label{eq.ham}
\ee
In this frame $\bk$ is constant, whereas $\bn$ varies due to the
evolution of the inclination. We have
\be
\bk=\bpm 0 \\ 0 \\ 1 \epm, \quad
\bn=\bpm -\sin i \\ 0 \\ \cos i\epm, \quad{\text{and}}\quad
\bw=\bpm x \\ y \\ z \epm
\ee
with $x=\sin\epsilon\cos\psi$, $y=\sin\epsilon\sin\psi$, and
$z=\cos\epsilon$. One can go from these variables to \citet{Ward_Hamilton_AJ_2004}
ones by a rotation of angle $i$ around the second axis. The equations of
motion associated to the Hamiltonian (\ref{eq.ham}) are given by
\be
\frac{d\bw}{dt} = \boldsymbol{\nabla}_{\bw}H\times\bw
\ee
which gives
\be
\frac{d\bw}{dt} = -\alpha(\dop{\bn}{\bw})\bn\times\bw-\nu{\bk}\times{\bw}
\ee
or equivalently
\be
\left\{
\EQM{
\dot{x} &=& \omega_zy, \cr
\dot{y} &=&-\omega_zx - \omega_x z, \cr
\dot{z} &=& \omega_xy,
}
\right.
\label{eq.motion}
\ee
with
\be
\EQM{
\omega_z &=& \alpha\cos i(z\cos i-x\sin i)+\nu, \cr
\omega_x &=& \alpha\sin i(z\cos i-x\sin i).
\label{eq.omega}
}
\ee
These equations are a combination of two rotations. The first one is a
rotation around the third axis with the angular velocity $\omega_z$.
The second is a rotation around the first axis
with the angular velocity $\omega_x$.

Now we describe the evolution of Uranus' spin-axis displayed in the
figure \ref{Fige}.

 During phase \one, from $t=0$ to $t=289$ kyr, the inclination $i$
and the obliquity $\epsilon$ are small, the three axes $\bw$, $\bn$, and
$\bk$ are almost aligned, there is no evolution.

 In phase \two, the variable $\psi$ remains close to $-\pi/2$ (Fig.
\ref{Fige}{\bf a}), there is thus a $1\!:\!1$ spin-orbit resonance.  In
that case, the angular velocity $\omega_z$ in (\ref{eq.motion}-\ref{eq.omega}) is
negligible and it remains only the rotation around the first axis
\be
\left\{
\EQM{
\dot{x} &\approx& 0, \cr
\dot{y} &\approx& -\omega_xz, \cr
\dot{z} &\approx& \omega_xy.
}
\right.
\label{eq.res}
\ee
As the axis $\bw$ describes an arc of a circle in the $y$-$z$ plane
(Fig. \ref{Fige}{\bf d}), the third component $z$ decreases, and thus the
obliquity $\epsilon$ increases (Fig. \ref{Fige}{\bf b}).

 In phase \three, the angle $\psi$ evolves (Fig. \ref{Fige}{\bf a}),
hence the resonance is broken and the angular velocity $\omega_z$
becomes important. On the other side, the inclination decreases (Fig.
\ref{Fige}{\bf c}). Thus, the angular velocity $\omega_x\propto\sin i$
decreases too.  Hence, the rotation in the $x$-$y$ plane dominates (Fig.
\ref{Fige}{\bf d})
\be
\left\{
\EQM{
\dot{x} &\approx& \omega_zy, \cr
\dot{y} &\approx& -\omega_zx, \cr
\dot{z} &\approx& 0,
}
\right.
\label{eq.nres}
\ee
\prep{\figd}
and the obliquity $\epsilon$ is almost stationary (Fig. \ref{Fige}{\bf
b}).

 During phase \four, the angle $\psi$ is stable for a second time.
The spin-axis is thus once again captured in resonance and the equations
of motion are the equations (\ref{eq.res}). The axis thus describes an
arc of a circle in the $y$-$z$ plane (Fig. \ref{Fige}{\bf d}). This time, it
starts with $y>0$ and $z<1$ ($\psi\approx\pi/2$) and goes toward $y=0$
and $z=1$.  As $z$ increases, the obliquity decreases (Fig.
\ref{Fige}{\bf b}). The orbital inclination $i$ is similar in phase
\two, and \four{} (Fig. \ref{Fige}{\bf c}), so is the angular velocity
$\omega_x$.

 When the spin-axis crosses the $x$-$z$ plane, the angle $\psi$ jumps
from $\pi/2$ to $-\pi/2$ and the system enters in phase \five{}
(Fig.  \ref{Fige}{\bf a}). The evolution is similar to the one of
phase \two.  The spin-axis describe an arc of a circle in the $y$-$z$
plane on the $y<0$ side (Fig. \ref{Fige}{\bf d}). The obliquity increases (Fig.
\ref{Fige}{\bf b}). However, as the inclination is higher than in
phase \two{} ($18\deg$, see Fig. \ref{Fige}{\bf c}), the angular
velocity $\omega_x$ is larger and the obliquity evolves faster (Fig.
\ref{Fige}{\bf b}).

 In phase \six, the resonance is once again broken (Fig.
\ref{Fige}{\bf a}), the angular velocity $\omega_x$ becomes negligible
with respect to $\omega_z$. The spin-axis describes an arc of a circle
in the $x$-$y$ plane (Fig. \ref{Fige}{\bf d}), and the obliquity remains
constant (Fig. \ref{Fige}{\bf b}).

After the tilt, the satellite is ejected. The precession constant
$\alpha$ decreases by a factor close to $1\,000$, which gives
$\omega_z\approx\nu$ and $\omega_x\approx0$. The equations of
motion (\ref{eq.motion}) thus show that the spin-axis precesses around
the third axis at the angular velocity $\nu$. The obliquity remains
constant as one can see in the figure \ref{Figb}.

\section{Conclusion}
We have shown that the current obliquity of Uranus is compatible
with planetary formation scenarios predicting small initial obliquities
without the need of a large collision. Moreover, we confirm the
necessity of the close encounters used in the Nice model to recover the
present eccentricity of Jupiter and Saturn. Additionally, we solve the
problem of the missing satellite around Uranus
\citep{Mosqueira_Estrada_Icarus_2003, Mosqueira_Estrada_Icarus_2003b}.
Although stellite formation theories are not at the stage that they can
constrain the final mass of the satellites, we acknoledge that the
satellite we have introduced may be too massive.
Nevertheless, a less massive satellite of only $0.001 M_U$ can still  be
sufficient to tilt Uranus if the planetary migration timescale is larger
than the one used here (see figure~\ref{Figd}). Several recent studies
actually suggest such a longer migration timescale
\citep{Murray-Clay_Chiang_ApJ_2005, Boue_etal_ApJL_2009,
Lykawka_etal_MNRAS_2009}. The parameters involved in the formation of
the Solar system are still too numerous to be able to derive precise
estimates of the probability of the present scenario to occur, but with
this additional satellite, we are able to propose a scenario that
fits with the present late migration scenario as given by the Nice
model. Depending on the progress made in satellite formation theories
and possible future variations in the migration scenarios, the elements
given here could provide strong additional constraints on the migration
timescales.

\acknowledgments
This work benefited from support from PNP-CNRS and CS from Paris
Observatory. The authors thank Paris Observatory SIO, and GENCI-CINES,
for providing the necessary computational resources for this work.


\prep{\small}
\prep{\renewcommand{\baselinestretch}{0.8}}

\bibliography{BL09}

\prep{\end{document}}


\clearpage

\figa

\figb

\figc

\clearpage

\figd

\fige


\end{document}